\documentclass[aps,prl,reprint,superscriptaddress,longbibliography]{revtex4-2}
\usepackage{graphicx}
\usepackage{amsmath}
\usepackage{mathcomp}
\usepackage{textcomp}
\usepackage{siunitx}
\usepackage{placeins}

\usepackage{xcolor}

\begin{document}


\title{Limits for coherent optical control of quantum emitters in layered materials}



\author{Michael K. Koch}
\email[M.K.K. and V.B. contributed equally to this work.]{}
\affiliation{Institute for Quantum Optics, Ulm University, D-89081 Ulm, Germany}
\affiliation{Center for Integrated Quantum Science and Technology (IQst), Ulm University, D-89081 Ulm, Germany}

\author{Vibhav Bharadwaj}
\email[M.K.K. and V.B. contributed equally to this work.]{}
\affiliation{Institute for Quantum Optics, Ulm University, D-89081 Ulm, Germany}
\affiliation{Department of Physics, Indian Institute of Technology Guwahati, 781039 Guwahati, Assam, India}

\author{Alexander Kubanek}
\affiliation{Institute for Quantum Optics, Ulm University, D-89081 Ulm, Germany}
\affiliation{Center for Integrated Quantum Science and Technology (IQst), Ulm University, D-89081 Ulm, Germany}


\date{\today}

\begin{abstract}
{The coherent control of a two-level system is among the most essential challenges in modern quantum optics. Understanding its fundamental limitations is crucial, also for the realization of next generation quantum devices. The quantum coherence of a two level system is fragile in particular, when the two levels are connected via an optical transition. When such quantum emitters are located in solids the coherence suffers from the interaction of the optical transition with the solid state environment, which requires the sample to be cooled to temperatures of a few Kelvin or below. Here, we use a mechanically isolated quantum emitter in hexagonal boron nitride to explore the individual mechanisms which affect the coherence of an optical transition under resonant drive. We operate the system at the threshold where the mechanical isolation collapses in order to study the onset and temperature-dependence of dephasing and independently of spectral diffusion. The new insights on the underlying physical decoherence mechanisms reveals a limit in temperature until which coherent driving of the system is possible. This study enables to increase the operation temperature of quantum devices, therefore reducing the need for cryogenic cooling.
}
\end{abstract}


\maketitle


Coherent control of quantum systems is a key requirement for upcoming quantum technologies. The conceptually simplest example is a two-level system which can be coherently manipulated by resonant driving leading to the discovery of coherent Rabi-oscillations \citep{Rabi39}. However, experimentally the coherence of a two-level system is fragile, for example, phase information can be lost rapidly. This becomes in particular evident, if both levels are connected by means of an optical transition where, for instance, the transition frequency could be sensitive to smallest disturbances such as electric or magnetic field fluctuations. The fragility becomes even more apparent if such a quantum emitter is embedded in a solid state matrix  \citep{Aharonovich16, Zhou18}, where for example electron-phonon interactions influence the optical coherence \citep{Kamada01, Skinner88}. Therefore, while optical control has emerged as a robust and fast way to manipulate quantum systems \citep{Wu10} it still suffers from fundamental limitations due to the decoherence of the quantum system. A study on the effects of the decoherence mechanisms will enable to develop improved control techniques, tailored to the specific quantum system.
Dephasing processes and the associated dephasing time describe the characteristic lifetime of a coherent polarization. It defines the time frame in which coherent manipulations can be performed and observed. In this work, we define the time required to apply an optical $\frac{\pi}{2}$-pulse as the minimum coherence time threshold to enable coherent optical control of a two-level quantum system. A $\frac{\pi}{2}$-pulse prepares the system in a coherent superposition between the ground- and excited state thereby creating coherence. From there on, dedicated pulse sequences, originally developed in the framework of nuclear magnetic resonance imaging \citep{Mamin07}, can be applied to extend the coherence times for specific applications. As an example, Ramsey-pulse sequences  \citep{Ramsey51} led to the development of atomic clocks, applied to clock transitions in the microwave regime, which has turned into an important application based on the coherence of quantum systems \citep{Ramsey83,Burt21,King22}.\\
In this work, we analyze the coherence of an optical transition of a single quantum emitter in a solid state matrix under resonant drive and study the mechanisms contributing to the decoherence of the system. We separately probe the influence of spectral diffusion and dephasing on the coherence and study their temperature-dependence. Defect centers in two-dimensional materials are interesting candidates for this study, since the layered structure of the host material provides the potential for mechanical isolation of the quantum emitters orbitals from phonon modes \citep{Hoese20}. In particular, defect center in hexagonal boron nitride (hBN) have recently emerged as a special quantum system, hosting quantum emitters with optical coherence \citep{Sontheimer17} retained up to room temperature \citep{Dietrich20}, which was inferred by studying the spectral linewidth of the optical transition under resonant excitation \citep{Dietrich18}. Here, we study the temperature dependent effect of both, dephasing and spectral diffusion, on the optical coherence, in particular, on coherent optical Rabi driving of quantum emitters in hBN. The temperature-dependence over a large temperature range gives new experimental input for the development of consistent models for the mechanical isolation. Unraveling the limiting factors allows coherent optical control to be brought to operation temperatures up to room temperature impacting upcoming quantum technologies.

\section{Separation of Individual Decoherence Mechanisms}
\begin{figure}[htbp]
\centering\includegraphics[scale=1]{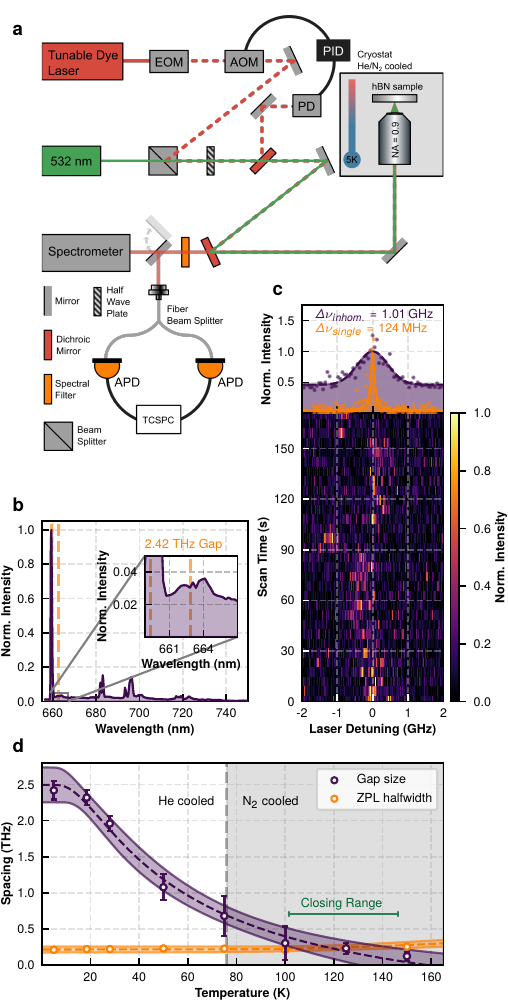}
\caption{\label{fig:overview} \textbf{Properties of the solid state quantum system. a} Schematic depiction of 
the optical setup. The flow cryostat 
allows the use of liquid 
helium for temperatures between $\SI{5}{\K}$ and $\SI{76}{\K}$ or liquid nitrogen for 
higher temperatures. The abbreviations used, 
but not explained in text are EOM (Electro-Optic Modulator), 
AOM (Acousto-Optic Modulator), PID 
(proportional-,integral- and differential 
regulator), PD 
(photo diode), APD (avalanche photodiode) and 
TCSPC (time-correlated single photon counter). \textbf{b} A PL spectrum of the emitter 
at $\SI{5}{\K}$ and a zoom-in on the first acoustic 
phonon mode indicating the gap with the ZPL. The difference between the 
spectral position of the ZPL and the displayed 
phonon mode span an energy gap of $\SI{2.42}{\tera
\hertz}$. \textbf{c} Long term PLE scan over the ZPL of the emitter displaying blinking, spectral diffusion and jumps at $\SI{5}{\kelvin}$. The homogeneous and 
inhomogeneous linewidth are shown on top. \textbf{d} Temperature-dependence of the energy gap and the ZPL half width. The purple and orange shaded area correspond to the $2 \sigma$-confidence interval of the respective fits. The green bar indicates the temperature regime where the closing of the gap occurs.}
\end{figure}

hBN is a promising host for bright quantum emitters with high stability, large Debye-Waller factor and polarized photon emission \citep{Kubanek22, Hoese22, Vaidya23, Malein21, Gupta23}, promising spin properties \citep{Gottscholl21, Ramsay23, Liu22} and ultrafast coherent state manipulation \citep{Preuss22}. We used one such quantum emitter in a hBN flake (methods) for our study. The schematic of the setup used for characterizing the emitter is shown in 
Fig. \ref{fig:overview}a. The photoluminescence (PL) spectra of the emitter at $\SI{5}{\kelvin}$ shows a narrow zero phonon line (ZPL) at around $\SI{660}{\nano\meter}$ (Fig. \ref{fig:overview}b) with a linearly polarized  emission (supplementary S1). Photoluminescence excitation (PLE) scans performed over the ZPL at $\SI{5}{\kelvin}$ (Fig. \ref{fig:overview}c) with a tunable laser shows that the emitter displays spectral diffusion indicated by the change in the position of the peak, spectral jumps in the position of the center of the peak and occasional dark state, with no peak seen. Some single scans $\left( \Delta \nu_{single} \right)$ over the ZPL show homogenous linewidths of $\SI{112 \pm 17}{\mega\hertz}$, which are fundamentally limited by the finite lifetime of the excited state without any further broadening, known as Fourier Transform Limited (FTL) linewidth ($\SI{109 \pm 9}{\mega\hertz}$) (supplementary S1). The 
inhomogenous linewidth $\left( \Delta \nu_{inhom.} \right)$ acquired over multiple scans shows a Gaussian distribution with a 
width of $\SI{1.01 \pm 0.07}{\giga\hertz}$, which is about 9 times broader than the FTL linewidth, indicating an improvement of the spectral stability of this class of emitters at cryogenic temperatures \citep{Hoese20}. The observation of lifetime limited emission lines is attributed to the presence of an energy gap between the ZPL and the first acoustic phonon mode (Fig. \ref{fig:overview}b inset). This gap represents a decoupling of the electronic orbital from low energy phonon modes and hence provides robust, isolated optical transitions enabling coherent optical manipulation. With increase in temperature, coupling to phonon modes increases, which leads to a closing of the gap. In our 
case, temperature-dependent PL spectra were acquired with liquid helium and liquid nitrogen cooling. The gap size and ZPL halfwidth were acquired from each of the PL spectra (supplementary S1). They were plotted as a function of temperature (Fig. \ref{fig:overview}d) and fitted with the Boltzmann function \citep{Hoese20}
\begin{align}
w = A + B \cdot \exp{\left(-\frac{C}{\mathrm{k_B} T}\right)} \, ,
\label{eq_boltzmann}
\end{align}
where $w$ describes both the temperature dependence on the gap size and the ZPL half-width. The variables $A$, $B$, and $C$ are free fit parameters. Here, we considered the intersection of the confidence intervals of the two fits as the closing range of the gap, which approximately was determined as $100 - \SI{150}{\kelvin}$ (Fig. \ref{fig:overview}d). Previously, FTL emission lines have been observed upto the temperature at which the gap size is about half of the value at $\SI{5}{\kelvin}$ \citep{Hoese20}. Hence, our temperature-dependent PL measurements provided a rough estimate of the temperature regime to study the spectral properties of resonant emission before and after the electronic orbitals start to couple to the phonon modes.

\section{Thermal induced spectral diffusion}

\begin{figure}[htbp]
\centering\includegraphics[scale=1]{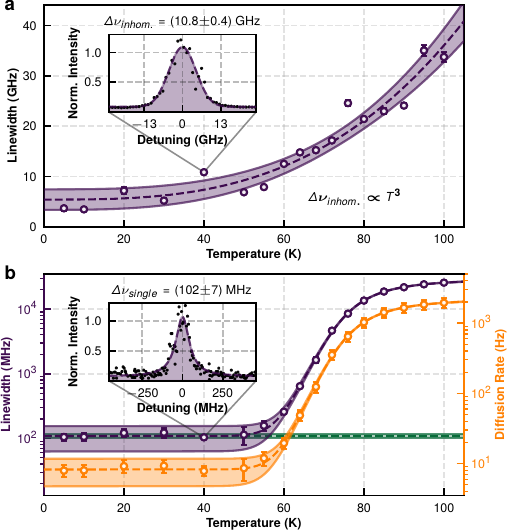}
\caption{\label{fig:ple_study} \textbf{Spectral properties at resonant excitation. a} The inhomogeneous linewidth, obtained through PLE spectroscopy, plotted against temperature showing a $T^3$-dependence. The inset figure displays the exemplary measured line shape with a gaussian fit at $\SI{40}{\K}$. The shaded area represents the $2 \sigma$-confidence interval of the fit. \textbf{b} The single scan PLE linewidth, measured with fixed scan parameters. It remains within the FTL up to $\SI{50}{\K}$. The FTL is displayed by the green shaded area, also representing the standard deviation. Additionally, a lower bound for the diffusion rate of the emitter is presented by the orange data (right y-axis). The left and right y-axes are logarithmically scaled. Both shaded areas depict the $1 \sigma$-confidence intervals of the respective fits. The inset figure shows the lorentzian line shape of a single scan over the ZPL transition at $\SI{40}{\K}$, indicating FTL linewidth.}
\end{figure}

More details on the underlying mechanisms behind the onset of electron-phonon coupling can be obtained by studying the temperature-dependence of the optical linewidth under resonant excitation. The emitter was found to be stable and showed emission under resonant excitation even with liquid nitrogen cooling temperatures, indicating a technical advancement in cryogenic measurements. Temperature-dependent PLE scans, performed with fixed scan parameters (Methods), reveal an increase of the inhomogenous linewidth with a $T^3$-dependence (Fig. \ref{fig:ple_study}a) \citep{Dietrich20,Horder22}. The data is fitted with
\begin{align}
\Delta \nu_{inhom.} = A + B \cdot T^3 \, .
\label{eq_inhom_lw}
\end{align}
The temperature-independent linewidth $A$ together with the scaling factor $B$ describe the inhomogeneous linewidth $\Delta\nu_{inhom.}$. Both, $A$ and $B$ are free fit parameters. Single PLE scans over the ZPL show linewidths within the FTL for temperatures upto 50 K. Increasing the temperature further, leads to a drastic increase in $\Delta \nu_{single}$ (Fig. \ref{fig:ple_study}b),which can be correlated with the onset of electron-phonon coupling. We found that the temperature-dependence of $\Delta \nu_{single}$ can be described with the following expression 
\begin{align}
\Delta \nu_{single} = D + \frac{A - D}{\left[ 1 + \exp{ \left( B \left( \log{ \left( T \right)} - C \right) \right)}\right]^E} \, .
\label{eq_logistic_regression}
\end{align}
The parameters $A$ and $D$ describe a lower and upper 
asymptote, which in our case translates to the FTL 
linewidth and the saturation of the observed linewidth for 
a single scan, respectively. $B$ ($B = \SI{20.0 \pm 1.0}{}$) is the slope parameter and 
describes the polynomial order at which the linewidth 
increases. The asymmetry of the curve is dictated by parameter 
$E$. Together with $B$ and $E$, the parameter $C$ 
encapsulates the halfway point between both asymptotes. 
Therefore, these parameters also describe the point when $
\Delta \nu_{single}$ crosses the FTL. With increase in 
temperature, the linewidth for a single scan approaches the 
full inhomogeneous linewidth of the emitter (Fig. 
\ref{fig:ple_study}a) and therefore saturates. By 
considering the laser scan speed $u_L$, the natural 
linewidth $\Delta \nu_{FTL}$ and the linewidth of a single 
scan $\Delta \nu_{single}$, a lower limit for the diffusion 
rate at a given temperature can be extracted from the PLE 
scans as $\left(u_L/\Delta \nu_{FTL} \right) \cdot 
\left(\Delta \nu_{single}/ \Delta \nu_{FTL} \right)$ 
\citep{Tran18}. Up to $\SI{50}{\kelvin}$ a constant 
diffusion rate of $\SI{8.5 \pm 1.9}{\hertz}$ was observed. A further increase in temperature leads to a sharp increase in diffusion rate to several kilohertz (Fig. \ref{fig:ple_study}b) (supplementary S2). Due to the linear relationship with $\Delta \nu_{single}$, the temperature-dependence of the diffusion rate was also fitted with equation \eqref{eq_logistic_regression} and therefore exhibits the same power dependence. It can be noted, that at 
low temperature, the diffusion rate is slow and the 
homogeneous linewidth is within the FTL. Hence,  we expect 
to observe coherent driving of the system in this regime.

\section{Coherent driving in the regime of low spectral diffusion}
\begin{figure}[htbp]
\centering\includegraphics[scale=1]{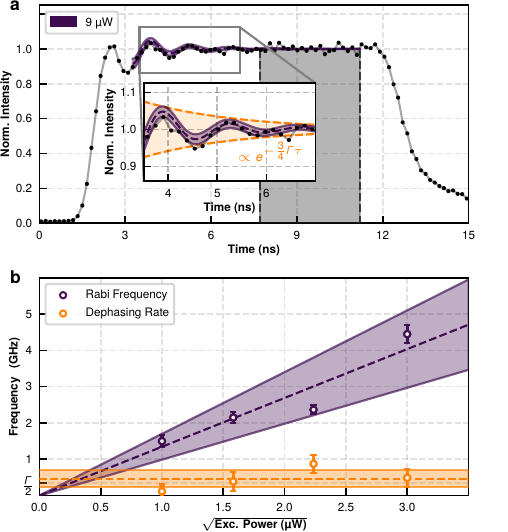}
\caption{\label{fig:pulsed_rabi} \textbf{Direct observation of optical Rabi oscillations. a} The resonant pulsed signal showing Rabi oscillations (fitted with eq. \eqref{eq_diffusion_g2}) when the optical transition is resonantly excited with a $\SI{10}{\nano\second}$ laser pulse with a power of $\SI{9}{\micro\watt}$. A zoom-in of the oscillating signal is shown in the inset. The purple shaded areas display the $2 \sigma$-confidence intervals of the respective fits and the gray shaded area represents the normalization range of the pulse. The envelope of the time evolution is described by an exponential decay, which only depends on the fluorescence decay rate $\Gamma$ \textbf{b} Fitted Rabi frequencies and dephasing rates at excitation laser powers of $\SI{1}{\micro\watt}$, $\SI{2.5}{\micro\watt}$, $\SI{5}{\micro\watt}$ and $\SI{9}{\micro\watt}$ as a function of the square root of the resonant excitation laser power. Both shaded areas represent the $2 \sigma$-confidence intervals of the respective fits.
}
\end{figure}

The population of the ground and excited state of a two-level system (TLS) oscillates when the system is coherently driven with sufficient driving power. The frequency of the oscillation, known as Rabi frequency, dependents on the excitation power and the dipole moment of the transition. The observance of such Rabi oscillations is the essential step to establish coherent optical control of the quantum system \citep{Allen87}. Studying the temperature-dependent decay of the Rabi oscillations provides precise insights into the decoherence mechanisms of the system. At 5K, direct optical Rabi oscillations were observed by pulsing 
the resonant excitation laser (Methods) with powers above saturation power ($P_{sat}$).
Oscillations in the population of the excited state become apparent for an excitation power of $\SI{9}{\micro\watt}$ (Fig. \ref{fig:pulsed_rabi}a). Rabi oscillations on an optical pulse were measured with different excitation powers (Fig. 3b)(supplementary S3). Observing coherent Rabi oscillations is the prerequisite to establish coherent control of a TLS by means of optical $\pi$- and $\frac{\pi}{2}$-pulses. However, in order to gain further understanding of the underlying limitations for coherence we now discuss the photon statistics by inferring second-order autocorrelations $g^{\left(2\right)}(\tau )$.

The temporal dynamics of the excited state population at time $\tau$ can be described by
\begin{align}
g^{\left(2\right)}\left(\tau\right) \propto \int_{-\infty}^{\infty} C\left(\Delta\right)^2  g^{\left(2\right)}\left(\tau , \Delta\right) \mathrm{d} \Delta \, ,
\label{eq_diffusion_g2}
\end{align}
(supplementary S5). Here, $C\left( \Delta \right)$ is proportional to the excited state population and depends on the detuning $\Delta$ from resonance. Therefore, $C\left( \Delta \right)$ describes the photon emission rate \citep{Loudon00, Flagg09, Fournier23}. Because the $g^{\left(2\right)}$ is measured as a two photon coincidence, the photon emission rate has a quadratic weighting in the integral. $g^{\left(2\right)}\left(\tau , \Delta\right)$ is the modified second-order autocorrelation function in presence of a fixed finite detuning from resonance. Hence, eq. \eqref{eq_diffusion_g2} describes the complete temporal dynamic of the excited state population, because it considers all possible spectral detunings and therefore also the effect of spectral diffusion on the optical coherence. 
The following equation describes the $g^{\left(2\right)}$-function exactly on resonance, accounting for both radiative decay and transverse dephasing \citep{Wrigge08}:
\begin{align}
g^{\left(2\right)}\left(\tau\right) = 1 + \frac{\lambda_-}{2 q} \exp{\left( \left| \tau \right| \lambda_+ \right)} - \frac{\lambda_+}{2 q} \exp{\left( \left| \tau \right| \lambda_- \right)} \, ,
\label{eq_resonant_g2}
\end{align}
where the substitutes $\lambda_\pm$ and $q$ are given by
\begin{align}
\lambda_{\pm} = -\frac{\Gamma + \Gamma_\perp}{2} \pm q \, ; \quad q = i \sqrt{\Omega^2 - \left(\frac{\Gamma - \Gamma_\perp}{2}\right)^2} \, .
\end{align}
$\Gamma$ is the fluorescence decay rate and $\Gamma_\perp$ the transverse dephasing rate 
\begin{align}
\Gamma_\perp = \frac{\Gamma}{2} + \gamma_c \, .
\label{eq_gamma_transverse}
\end{align}
Generally, $\gamma_c$ is referred to as the pure dephasing rate \citep{Zhou17,Skinner86} and includes the temperature- and power-dependence in our observations, whereas $\Gamma$ is constant. In an atom cloud, pure dephasing can be caused by atom-atom collisions. Whereas in a solid state crystal, phonon interactions are a main cause for pure dephasing \cite{Grange17, Geng23, Skinner88}. Since these types of interactions do not affect the excited state population, the coherence can be recovered up to the limit given by the fluorescence decay rate. 
For optical driving detuned from resonance, the Rabi frequency $\Omega$ takes on the general form $\sqrt{\Omega^2 + \Delta^2}$, which makes $g^{\left(2\right)}\left(\tau\right)$ in eq. \eqref{eq_resonant_g2} also dependent on $\Delta$. 

The direct optical Rabi oscillations (Fig. \ref{fig:pulsed_rabi}a) were fitted with eq. \eqref{eq_diffusion_g2}, which accounts for the average detuning during the measurement time. Both, $\Omega$ and $\Gamma_\perp$ are free fit parameters. The reduced contrast (Fig. \ref{fig:pulsed_rabi}a) is attributed 
to spectral diffusion, because the fluorescence
\begin{align}
C\left( \Delta \right) \propto \frac{1}{2} \frac{\Omega^2 \Gamma_\perp / \Gamma}{\Delta^2 + \Gamma_\perp^2 + \Omega^2 \Gamma_\perp / \Gamma} \, ,
\label{eq_excitation_prob}
\end{align}
monotonically decreases for increasing detuning. The signal to noise ratio is also hampered by the background florescence signal from the substrate. The signature power dependence of the Rabi frequency and a constant transverse dephasing rate (Fig. \ref{fig:pulsed_rabi}b) confirms coherent optical driving of a two level system. The constant dephasing rate at $\SI{5}{\kelvin}$ allows us to observe coherent optical driving at any power above saturation.

\section{Thermal induced dephasing}

\begin{figure*}[htbp]
\centering\includegraphics[scale=1.0]{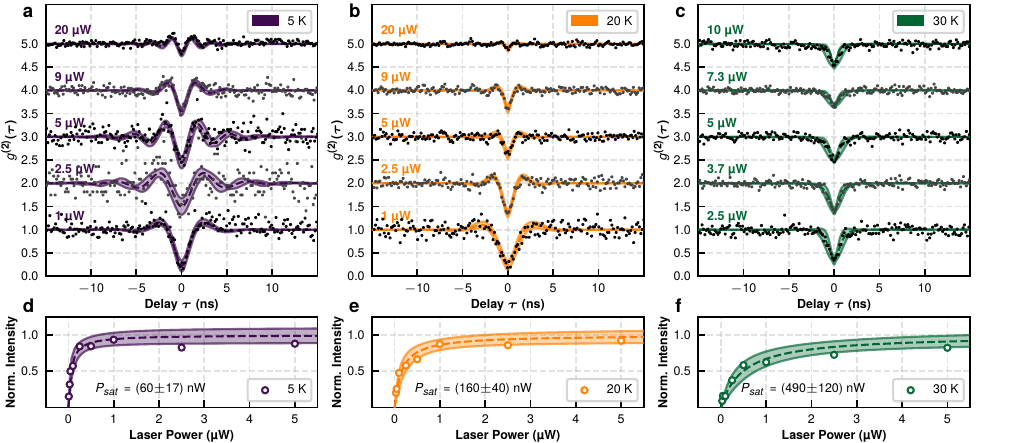}
\caption{\label{fig:rabi_g2} \textbf{Photon correlation measurements. a-c} The raw data points (black and gray dots) for three different temperatures ($\SI{5}{\K}$, $\SI{20}{\K}$ and $\SI{30}{\K}$) of the second-order correlation measurement at resonance with various powers. Each measurement is fitted with eq.\eqref{eq_diffusion_g2}, normalized to 1 and offset by 1 for each consecutive power. \textbf{d-f} Intensity saturation measurements for the temperatures depicted in part \textbf{a-c}. The saturation parameters are fitted with eq.\eqref{eq_int_sat}. All shaded areas in part \textbf{a-f} show the $2 \sigma$-confidence interval of the respective fits.}
\end{figure*}

To understand the effect of increasing temperature on the dephasing rate, we investigated the photon statistics at multiple temperatures and powers above $P_{sat}$. The photon statistics is inferred from second-order photon correlations (Methods) under resonant, continous wave (CW) laser excitation in Hanbury Brown-Twiss (HBT) configuration as shown in Fig. \ref{fig:overview}a. The second-order correlations are measured at $\SI{5}
{\kelvin}$, $\SI{20}{\kelvin}$ and $\SI{30}{\kelvin}$ (Fig. \ref{fig:rabi_g2}a-c) with powers above $P_{sat}$ for each of the temperatures (Fig. \ref{fig:rabi_g2}d-f). At each of the temperatures, the intensity saturation powers were extracted from
\begin{align}
I\left(P\right) = I_{\infty} \frac{P/P_{sat}}{1 + P/P_{sat}} \, ,
\label{eq_int_sat}
\end{align}
where $I\left(P\right)$ is the excitation power ($P$) dependent emission intensity of the emitter and $\frac{I_{\infty}}{2}$ is the saturation intensity. Each correlation measurement was acquired for about $\SI{9}{\hour}$ with continuous resonant excitation on the emitter. Unlike the direct Rabi detection on an optical pulse, these long measurements provide a complete understanding of all effects which play a role in the decoherence mechanisms and are not influenced by the non-ideal shape of a resonant laser pulse. Decoherence mechanisms include, in particular, all effects resulting from the observed spectral instabilities and dephasing. A signature for the Rabi oscillations are the appearance of shoulders 
in the $g^{\left(2\right)}(\tau )$-function next to the dip at zero time delay \citep{Horder22, Konthasinghe19}. The shoulders indicate the onset of Rabi oscillations and were observed at $\SI{5}{\kelvin}$ and $\SI{20}{\kelvin}$. At $\SI{30} {\kelvin}$, there is no clear appearance of the oscillations on the shoulder, but the width of the dip becomes narrower with increasing power. The data is fitted with eq.\eqref{eq_diffusion_g2}. For the data set measured at $
\SI{5}{\kelvin}$ and $\SI{20}{\kelvin}$ the analysis shows that $\lambda_\pm$ has an imaginary part. This means that the dephasing rate is lower than the Rabi frequency. Therefore, the excited state population oscillates sinusoidal. At $\SI{30}{\kelvin}$ however, $\lambda_
\pm$ only have a real part for any power and therefore the observed dynamics can be described as an overdamped oscillation. Consequently, the dephasing rate is always larger than the associated Rabi frequency.
  
\begin{figure}[htbp]
\centering\includegraphics[scale=1]{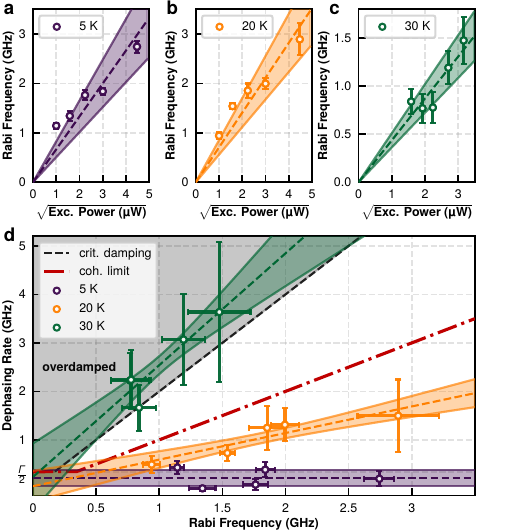}
\caption{\label{fig:rabi_vs_dephasing} \textbf{Limits of coherent optical driving. a-c} Fitted Rabi frequencies of the second-order correlation measurements as a function of square root of the incident laser power from fig. \ref{fig:rabi_g2}. The shaded areas correspond to the $2 \sigma$-confidence interval of the respective fit. \textbf{d} Transverse dephasing rate $\Gamma_\perp$ plotted against the associated Rabi frequency for different temperatures. The gray shaded area describes the case of an overdamped oscillation and the black dashed line defines the case of critical damping (slope of 2). A slope of 1 defines the coherent driving limit, which is depicted by the red dotted-dashed line. Increasing Rabi frequency indicates increasing excitation power. The purple shaded area shows the $2 \sigma$-confidence interval, whereas, for clarity, the orange and green shaded area displays the $1 \sigma$-confidence interval.}
\end{figure}

The power dependence of the acquired Rabi frequency 
for $\SI{5}{\kelvin}$, $\SI{20}{\kelvin}$ and $\SI{30}{\kelvin}$ are shown in Fig. \ref{fig:rabi_vs_dephasing}a-c. The linear dependence of the Rabi frequency with the square root of the incident laser power is a clear confirmation for the observation of optical Rabi flopping. Further resonant auto correlation studies were performed for higher temperatures, also with liquid nitrogen (supplementary S4), again confirming the single photon nature of the emitter while coherent Rabi-oscillations are absent at higher temperatures. 

This study allows us to extract the transverse dephasing rates $\Gamma_\perp$ for varying temperatures, which helps to investigate the limits of optical coherent driving. The analysis shows a linear dependence on the square root of the excitation power for both, the Rabi frequency and the dephasing rate. Hence, the Rabi frequency can be studied depending on the dephasing rate (Fig. \ref{fig:rabi_vs_dephasing}d). Increasing the Rabi frequency refers to an increase in excitation power (Fig. \ref{fig:rabi_vs_dephasing}d). We use both, the slope and the offset of a linear fitting curve as free fitting parameters. The offset for all data sets agrees with the fundamental limit of $2 \Gamma_{\perp, min.} = \Gamma$, which is given by eq. \eqref{eq_gamma_transverse}. This rate provides the dark time decay rate of the coherence in the 
system, because it is the extrapolated value at zero power. Since there is a linear relation between $ \Gamma_\perp$ and $\Omega$, we can reduce the discussion about the limit of coherent driving to the slope of the associated linear fit. In case the dephasing rate is equal to the Rabi frequency at any given power, the longest pulse one can achieve is a $\frac{\pi}{2}$-pulse or, in other words, the coherence time is just long enough to prepare the system into a coherent superposition between ground- and excited state. Therefore, we consider a slope of one to define the ultimate limit for coherent manipulation. Additionally, for a slope with $m \leq 0.5$, it is possible to apply a $\pi$-pulse, or two $\frac{\pi}{2}$-pulses sandwiching a dark time, which enables dynamical decoupling techniques \citep{Viola99, Holzapfel20, Biercuk09, Du09} to recover full coherence up to the lifetime limit. At $\SI{5}{\kelvin}$, we observed a constant dephasing rate of $\SI{0.22 \pm 0.17}{\giga \hertz}$ and therefore a slope of zero. For $\SI{20}{\kelvin}$, the dephasing increases with the Rabi frequency. The slope is $\SI{0.55 \pm 0.14}{}$ and therefore below the limit for coherent driving, which means coherent manipulation of the system is possible at any point above the saturation power. The evaluated slope of $\SI{2.3 \pm 0.6}{}$ extracted from the measurement at $\SI{30}{\kelvin}$ is above the coherent driving limit and also above the limit for critical damping. At critical damping the dephasing rate is twice the Rabi frequency. Hence, it is described by a slope of $m=2$. Any slope above that represents an overdamped oscillation and is therefore described by a pure exponential decay. From our study, for this particular emitter, the temperature upto which coherent driving is achievable is between $\SI{20}{\kelvin}$ and $\SI{30}{\kelvin}$. 
With our observations, we assume the temperature-dependent coherence properties of a mechanical decoupled hBN defect would improve for an emitter where the closing point of the associated energy gap between the ZPL and the first acoustic phonon mode shifts to higher temperatures. Therefore, even if the phonon population increases with increasing temperature, the emitter retains its coherence properties due to decoupling from low energy phonon modes.

\section{Conclusion \& Outlook}
Summarizing, we study the limits for coherent, resonant driving of a single optical transition for a quantum emitter in a layered host material, in particular mechanically isolated defect center in hBN. We propose to define the time required to apply a $\frac{\pi}{2}$-pulse, which enables to create a coherent superposition between ground- and excited state, as the minimum time limit to claim coherence for the two-level system. We extract the temperature-dependence of the limiting factors, in particular, spectral diffusion and dephasing and study the results in context with the mechanical isolation of the defect center from low-frequency phonons. From this study, we understand that the primary contribution to the broadening mechanism is from the fast spectral diffusion at higher temperatures. Electric field tuning via Stark effect and surface charge passivation can help to reduce spectral diffusion \citep{Akbari22, Akbari21}. Furthermore, integration in optical devices \citep{Turunen22, Kianinia22}, such as cavities, will enhance emission rates and potentially improve the indistinguishability of photon emission \citep{Flagg09, Vogl19, Hausler21, Moon23, Froch22} due to spectral broadening originating from the Purcell effect \citep{Karanikolas23}. In particular for mechanically decoupled emitters in hBN, this study helps in developing consistent models to understand the mechanisms behind the decoherence \citep{Sharman23, Medeiros23}. In the future, mechanically isolated quantum emitters in hBN could enable coherent optical control at room temperature with impact on upcoming quantum technology.

\newpage
\section*{Methods}
\subsection{Sample Preparation}
The hBN sample was prepared with commercially available hBN micro-particles (2D Semiconductors) dissolved in ethanol. We applied the particles onto a sapphire substrate via spin coating. Prior to spin coating the substrate was acid boiled and plasma cleaned. The sample was then annealed in vacuum at $\SI{800}{\celsius}$ for $\SI{1}{\hour}$.
\subsection{Temperature-Dependent Photoluminescence}
The experimental setup is shown in Fig. \ref{fig:overview}a. The sample was placed in a vacuum sealed flow cryostat with the ability to be cooled by liquid helium or liquid nitrogen and has a PID controlled heater for setting a particular temperature. The system rested for 15 
minutes at each temperature to stabilize any temperature fluctuations. A high NA (NA=0.9) objective enables efficient excitation and collection of the emission from the hBN defect. For acquiring PL spectra, an off-resonant $\SI{532}{\nano\meter}$ laser was used to excite the emitter and the emission was collected onto the spectrometer after filtering the excitation laser with a $\SI{580}{\nano\meter}$ long-pass spectral filter. The gap size was measured by considering the spacing between the Gaussian fit for the ZPL and the low energy phonon mode. PL spectra for each temperature was measured with fixed green power of $\SI{80}{\micro\watt}$ and exposure time of $\SI{60}{\second}$. A pulsed green laser was used to measure the lifetime decay (supplementary S1) at various temperatures with fixed average laser power of $\SI{10}{\micro\watt}$ and exposure time of $\SI{15}{\minute}$ to obtain the lifetime of the emitter.
\subsection{Temperature-Dependent Photoluminescence Excitation}
For PLE spectroscopy, a tunable Dye laser was used to excite the emitter resonantly and the signal from the phonon side band was collected after filtering the ZPL spectrally with a tunable long pass filter. Fiber coupled APDs were used to collect emitted photons. The optical power from the laser 
was actively stabilized by a home built PID controller feedback loop connected to the AOM. To obtain the PLE scans at various temperatures, the laser was scanned at a rate of $\SI{890}{\mega\hertz\per\second}$ over the ZPL for 10 scans in both directions at a fixed resonant laser power of $\SI{10}{\nano\watt}$. The inhomogenous linewidth was obtained by fitting the histogram of the multiple scans with a Gaussian function. Individual scans up to a temperature of $\SI{50}{\kelvin}$ were analyzed using a Lorentz function, which agree with the FTL. Above $\SI{55}{\kelvin}$, a Gaussian profile was used for analysis due to the observed broadening of the linewidth. 
\subsection{Rabi Flopping on a Bare Pulse}
The direct detection of Rabi was measured by pulsing the resonant laser using an EOM and an FPGA pulser. The pulse signal was acquired with a $\SI{160}{\pico\second}$ resolution, after $\SI{5}{\minute}$ of measurement with a sequence consisting of a $\SI{10}{\nano\second}$ resonant pulse and $\SI{1}{\micro\second}$ of break time. The bare pulse showed a finite rise time of $\SI{1.3}{\nano\second}$ (supplementary S3).
\subsection{Photon Correlation Measurements}
Hanbury Brown-Twiss configuration was used for photon correlation measurements, wherein the emitted photons passed through a 50:50 fiber beam splitter connected to TCSPC (Fig. \ref{fig:overview}a). PID feedback controllers actively stabilized the resonant laser power. The resolution of the data acquisition was set to $\SI{160}{\pico\second}$. The correlation measurements for each temperature and power were acquired for about $\SI{9}{\hour}$. 
\medskip
\begin{acknowledgments}
The project was funded by the German Federal Ministry of Education and Research within the research program Quantum Systems in project 13N16741. 
A.K. acknowledges support of the Baden-Wuerttemberg Stiftung gGmbH in Project No. BWST-ISF2022-026. M.K.K. and A.K. acknowledge support of IQst.
V.B. acknowledges the support of the Alexander von Humboldt Foundation.
\end{acknowledgments}

\bibliographystyle{nature}
\bibliography{CohRabi.bib}

\end{document}